\documentclass[final,5p,times,twocolumn]{elsarticle}
\usepackage{ifpdf}
\usepackage{graphicx}
\usepackage{amssymb}
\usepackage{amsmath}
\usepackage{color}
\usepackage{xcolor}
\usepackage{multirow}
\usepackage{afterpage}
\usepackage{hyperref}
\journal{Physics Letters A}
\begin{document}
\begin{frontmatter}
\title{
Domain formation of modulation instability in spin-orbit-Rabi coupled
Gross-Pitaevskii equation with cubic-quintic interactions}
\author{
R. Sasireka$^1$, S. Sabari$^2$, and A. Uthayakumar$^1$, and Lauro Tomio$^2$
}
\address{$^1$Department of Physics, Presidency College (Autonomous), Chennai - 600005, 
India \\ $^2$Instituto de F\'\i sica Te\'{o}rica, UNESP -- Universidade Estadual Paulista, 
01140-070 S\~{a}o Paulo, Brazil}
\begin{abstract} 
The effect of two- and three-body interactions on the modulation instability (MI) domain 
formation of a spin-orbit (SO) and Rabi-coupled Bose-Einstein condensate is studied within 
a quasi-one-dimensional model. To this aim, we perform numerical and analytical investigations 
of the associated dispersion relations derived from the corresponding coupled 
Gross-Pitaevskii equation. The interplay between the linear (SO and Rabi) couplings with 
the nonlinear cubic-quintic interactions are explored in the mixture, considering miscible 
and immiscible configurations, with a focus on the impact in the analysis of experimental 
realizations with general binary coupled systems, in which nonlinear interactions 
can be widely varied together with linear couplings. 
\end{abstract}
\end{frontmatter}
\date{\today}

\section{Introduction}
\label{sec:1}
{Modulational instability (MI) is a generic phenomenon leading to large-amplitude periodic waves, 
which occurs in dynamic systems, like fluids, nonlinear optics, and plasmas. It results from the 
interplay between nonlinear dynamics and dispersion (or diffraction in the spatial domain), 
with the fragmentation of carrier waves into trains of localized waves~\cite{1967Benjamin,1970Hasegawa}, 
corresponding to the growth of weakly-modulated continuous waves in a nonlinear medium. Considering 
experimental setups, it was also reported recently investigations in optics and
hydrodynamics~\cite{2021Vanderhaegen}, demonstrating that MI can be a more complex phenomenon than the 
one predicted by the conventional linear stability analysis~\cite{2020Conforti}, revealing that MI goes 
beyond the limited predicted frequency range.
{\color{black} In Bose-Einstein condensates (BECs), we have already the experimental realizations 
reported in Refs.~\cite{2017Nguyen,2017Everitt} of MI occurring in cigar-shaped trapped condensates,
which are demonstrating the relevance of MI in cold-atom physics.
In~\cite{2017Nguyen}, by varying the two-body interaction of a condensate with the lithium isotope $^7$Li,
the authors reported the formation of matter-wave soliton trains due to MI; whereas in 
\cite{2017Everitt} the authors have considered the Rubidium isotope $^{85}$Rb in their experiment,
showing that MI is the key underlying physical mechanism driving the breakup of the condensate.
}
Besides that, particularly when considering coupled channel systems, within higher-order nonlinear interaction 
effects, further investigations are still required to be pursued by using conventional MI linear analysis, 
in order to clarify the most relevant distinguished outcomes, which can possibly emerge in actual cold-atom 
experiments. By following some other investigations reported in 
Refs.~\cite{2002Khawaja,2002Strecker,2004Carr}, 
within the nonlinear Schr\"odinger (NLS) formalism, as the Gross-Pitaevskii (GP) equation, 
plenty of other studies considering MI analysis have been performed in the last two decades. 
Among them, a variational analysis was performed in Ref.~\cite{2007Ndzana,Sabari2020}, for cubic-quintic NLS. 
Other studies are, for example, by assuming scalar~\cite{2013Sabari,2015Sabari,2014Wamba,Sabari2022} 
and vector~\cite{Goldstein,Kasamatsu2004,Kasamatsu2006,Sabari2019} BECs.
As verified in~\cite{ZRaptiMI}, 
the above scenario can
drastically change when considering discrete multi-component systems.}
It has been found that the onset of MI not only depends on the sign of diffraction (or dispersion), 
but also on the sign of the inter- and intra-species interaction strengths. Beyond the GP mean-field formalism, 
by considering quantum fluctuations through the Lee-Huang-Yang (LHY) term~\cite{1957LHY}, the MI was studied
in Refs.~\cite{2019-Abdullaev,2022-Otajonov}, following analysis of Faraday wave patterns and droplets 
generated in Bose gas mixtures.  

By considering spin-orbit (SO) and Rabi couplings in BECs, with the usual assumption of two-body nonlinear 
interaction in the GP formalism, MI analysis has been recently performed in continuous~\cite{2015Bhat,2019Mithun}, and in discrete media~\cite{Sabari2021}. 
The interest in considering MI in BECs with SO and Rabi couplings follows 
the experimental realization of synthetic spin-orbit coupling (SOC), as described 
in Ref.~\cite{Lin2011}, in which two Raman laser beams were used to couple two-component BECs. 
The momentum transfer between the laser beams and the atoms leads to the synthetic SOCs~\cite{Ruseckas2005}, 
realized in cold atom gases by designating the hyperfine atomic states as pseudo-spins, which are coupled 
by Raman laser beams~\cite{Zhang2012}.

One should also notice the relevance of SO coupling studies, as associated with new developments in 
spintronic devices~\cite{Zutic,Nagaosa}, in topological quantum computation~\cite{Sau} and in hybrid
structures~\cite{Mishchenko,Avsar}. For instance, it was also reported that new ground-state phases can be 
created in two-component nonlinear systems, such as stripes and phase separations~\cite{Wang2010}, 
tricritical points~\cite{Li2012}, different soliton types~\cite{Achilleos,Kartashov,Lobanov,Salasnich2013}, 
two-dimensional (2D) solitons with embedded vorticity~\cite{Sakaguchi-Li-Malomed1,Salasnich-Malomed1,Sakaguchi-Meqs}. 
In Refs.~\cite{2014-Cheng, 2016-Salerno}, one can also follow some investigations on SOC by considering 
BEC in optical lattices. With tunable SOC and attractive interactions, under slow and rapid time Raman frequency 
modulations, Josephson oscillations were investigated in Ref.~\cite{2018-Abdullaev}, following previous 
related studies~\cite{levy2007,abbarchi2013}. More recently, the effect of SO and Rabi couplings on 
the excitation spectrum of binary BECs was investigated in Ref.~\cite{2021-RavisankarPRA}. In this work, 
by using the Bogoliubov-de Gennes (BdG) theory to explore the dynamical stability, it was verified the role of 
SO and Rabi couplings in the stability of a quasi-2D system, when considering repulsive contact interactions. 
As shown for the 2D system, the larger SOC increases the instability, whereas the increase of Rabi coupling 
favors stability. For cigar-type one-dimensional (1D) confining systems, some studies on the stability of 
SO-coupled systems are given in Ref.~\cite{Achilleos}. 

In spite of the advances evidenced in the above-described investigations, when considering SO and Rabi couplings 
as well as other more advanced studies going beyond the mean-field approach in BECs, such as quantum fluctuation 
effects due to the LHY term~\cite{1957LHY}, one can still follow some simple complementary studies which can be 
necessary for the quantitative analysis of experimental results concerning MI with SO and Rabi couplings in BECs.  
In this regard, the inclusion of quintic nonlinear effects in the NLS formalism, which can arise from triple 
particle collisions~\cite{2000Gammal,2000GammalJPB}, is expected to show up with some new parametric regions of 
instability, affecting previous modulation instability analysis.  As far as we know, 
higher-order nonlinear interaction effects in binary-coupled systems have not been explored in 
available works on MI instabilities, particularly when also considering linear couplings. Therefore, 
we understand this gap should be filled within analogies with previously related studies in which only 
two-body interactions are been taken into account, as the one provided Ref.~\cite{2015Bhat}. 
The relevance of this study relies on the fact that it has been documented that three-body interactions 
can bring significant contributions to studies on the properties and dynamics of BEC systems. 
Hence, an important question immediately arises on the unique role played by three-body interaction in the 
ground state properties and dynamics of SO-coupled BECs. Therefore, the main task we are considering with 
the present investigation is to verify how the inclusion of three-body interactions may impact the present 
knowledge related to the stability of SO-coupled BECs, which have been studied along with two-body interactions 
by using Rabi coupling. The freedom associated with the couplings (SO and Rabi), together with the intra- and 
inter-species interaction parameters are expected to lead to the identification of several interesting stable 
domains, which can be explored experimentally by considering different hyperfine states of the same kind of 
atoms. From this perspective, our work has a general purpose, 
with interactions and coupling parameters left to be fixed by the particular properties of coupled
atomic species being considered in experimental realizations.

Next, this paper has the following structure: In Sect.\ref{sec:2}, the basic formalism for the SOC model 
is presented in terms of a coupled GP equation. In Sect.~\ref{sec:3}, we provide details on the associated 
dispersion relations, which are in the fundamental framework of the MI analysis. The significant main results, 
obtained by using computational analysis together with analytical discussion, are provided in Sect.~\ref{sec:4}. 
Finally, in Sect.~\ref{sec:conclusion}, we resume and discuss our main conclusions. In a supplementary section, we have given numerical support for few cases.

\section{Model formalism}
\label{sec:2}
We consider two BEC atomic species coupled with SO interaction, trapped by harmonic potential, 
in which equal Rashba and Dresselhaus~\cite{1959Rashba,1955Dresselhaus} couplings are manifested, 
with the SO interaction causing the energy bands to split. We assume cigar-type confinement for the 
coupled system, with the perpendicular ($\omega_\perp$) and longitudinal ($\omega_x$) trap frequencies such 
that $\omega_\perp\gg\omega_x$. In the NLS coupled formalism, in addition to the main nonlinear cubic term
due to the two-body interactions, we also consider a quintic nonlinear term, which can arrive mainly due 
to three-body interactions. Besides being usually much smaller than the cubic term, in addition to other
small quantum fluctuations, this quintic term is expected to produce relevant effects in the 
dynamics of the coupled BEC system. our main purpose, in this regard, is to verify how it affects 
the MI analysis performed in previous studies.

In the following model approach, after reduction from the usual three-dimensional (3D) 
extended GP formalism, with the order parameters of the two pseudo-spin components ($i=1,2$), 
obtained from the corresponding field operators given by $\Psi_{1,2}(x,t)$, the corresponding 
1D formalism is described  as 
{\small
\begin{eqnarray}\label{eqGP0}
{\rm i}\hbar\frac{\partial}{\partial t} \left( \begin{array}{c} \Psi_1\\ \Psi_2 \end{array} \right)
& =& \left[H_0+ H_{NL}\right] \left( \begin{array}{c} \Psi_1\\ \Psi_2 \end{array} \right),\label{H} 
\end{eqnarray} 
}where $H_{NL}$ is the nonlinear part of the total Hamiltonian, with $H_0$ the linear part, 
carrying the SOC and Rabi frequency parameters.
Expressed with the usual Pauli spin matrices $\sigma_{x,z}$, $H_0$ is given by
\begin{eqnarray} H_0&\equiv & \frac{P_x^2}{2m} +\frac{\hbar\kappa}{m} P_x\sigma_z +
\hbar \frac{\Omega_R}{2}\sigma_x  + V(x) ,\label{H0} 
\end{eqnarray}
where $m$ is the atom mass, $\kappa$ is the SOC strength parameter for the two hyperfine states, and
$\Omega_R$ the Rabi frequency. 
The trapping potential in the $x-$direction, ${V}(x)$, will be assumed as zero, in our
case that $\omega_\perp\gg\omega_x$.
{\color{black}
In the recent experimental studies on modulational instability in BECs, 
in Ref.~\cite{2017Nguyen} using $^7$Li and in Ref.~\cite{2017Everitt} using $^{85}$Rb,
the authors have considered elongated cigar-type condensed clouds with 
$\omega_\perp\approx47\omega_x$ and $\omega_\perp\approx10\omega_x$, respectively,
which are indicating that our assumption is quite realistic.
}The $H_{NL}$ defines the nonlinear part of the Hamiltonian, with the 
two- and three-body interaction parameters:
{\small\begin{eqnarray}
H_{NL}&\equiv& \left(
\begin{array}{cc}
\sum_{j,n} g_{1j}^{(n)}|\Psi_j|^{2n}& \hspace{-0.5cm}0\\
0&\hspace{-0.5cm}\sum_{j,n} g_{2j}^{(n)}|\Psi_j|^{2n}
\end{array}
\right) \nonumber\\
&+& 2|\Psi_1|^2|\Psi_2|^2\left( \begin{array}{cc}
g_{12}^{(2)}&0\\ 0&g_{21}^{(2)}\end{array}\right)
\label{HNL}
.\end{eqnarray}
}where the level of nonlinearity is labeled by $n=1,2$.
For the cubic nonlinear interactions, $g_{ij}^{(1)}\equiv 2a_{ij}$ ($i,j=1,2$)
are the 1D reduction of the corresponding 3D parameters ($4\pi\hbar^2 a_{ij}/m$) associated to the
$s-$wave scattering lengths $a_{ij}$ between the $i,j$ components;
 whereas the 1D quintic nonlinear parameters, $g_{ij}^{(2)}$ are mainly associated to three-body 
 interactions.  In both cases, for cubic and quintic terms in the nonlinear Schr\"odinger equation,  
 we are assuming the symmetry $g_{12}^{(n)}=g_{21}^{(n)}$.
The three-body parameters effectively arise from two-body ones, as derived from effective field theory in 
\cite{1999Braaten}, and also reported in \cite{2013Jibbouri}.
However, one can also follow the model approach presented in Ref.~\cite{2021Hammond}, 
where they are indicating how the three-body parameter can be controlled independently from 
the two-body one.
In the above expressions \eqref{H}-\eqref{HNL}, the two components are assumed normalized to the 
total number $N$ of atoms, 
\begin{equation}
\Psi\equiv \left( \begin{array}{c} \Psi_1\\ \Psi_2 \end{array} \right),\;\;\;
\sum_{j=1}^2 \int dx |\Psi_j|^2 =N ,
\label{norm0}\end{equation}
with the functional energy given by
{\small\begin{align}
E[\Psi_{1},\Psi_{2}]&=\int d{x}\Big\{\left(\Psi ^{T} H_{0} \Psi \right)+
\frac{1}{2} \Big[{g}_{11}^{(1)}\lvert \Psi_{1} \rvert^{4}+{g}_{22}^{(1)} 
\lvert \Psi_{2} \rvert^{4}\notag \\
&+2{g}_{12}^{(1)} \lvert \Psi _{1} \rvert^{2} \lvert \Psi _{2} \rvert^{2}\Big] 
+\frac{1}{3}\Big[{g}_{11}^{(2)}\lvert \Psi_{1} \rvert^{6}+{g}_{22}^{(2)}\lvert 
\Psi_{2} \rvert^{6}
\notag \\
&+3{g}_{12}^{(2)}\lvert \Psi _{1} \rvert^{2} \lvert \Psi _{2} \rvert^{2}
\left(\lvert \Psi_{1} \rvert^{2}
+ \lvert \Psi _{2} \rvert^{2}\right) \Big]\Big\}.
\label{eq:e1d}
\end{align}
}Let us consider the above formalism in dimensionless quantities. For that, 
{\color{black} 
all the frequencies will be given as functions of the radial trap 
frequency $\omega _{\bot}$, with $\omega _{\bot}^{-1}$ being the time unit,
$\hbar \omega _{\bot }$ the energy unit, and  
$\ell_{\bot }=\sqrt{\hbar /(m\omega _{\bot})}$ the length unit.}
For the linear and nonlinear parameters, 
we redefine the 
 Rabi frequency as $\nu_R\equiv\Omega_R/(2\omega_\perp)$, the SOC strength as 
 $\gamma\equiv\kappa\sqrt{\frac{\hbar}{m\omega_\perp}}$, with the cubic and 
 quintic nonlinear parameters, respectively, given by $\bar{g}_{ij}
 \equiv\frac{g_{ij}^{(1)}}{\ell_\perp\hbar\omega_\perp}=\frac{2a_{ij}}
 {\ell_\perp\hbar\omega_\perp}$ and $\bar{\chi}_{ij}=\frac{g_{ij}^{(2)}}
 {\ell_\perp^2\hbar\omega_\perp}$.
Within these units and parameters, the two-component wave functions redefined as
$\psi_{j=1,2}\equiv \psi_{j}(x,t)\equiv\sqrt{\ell_\perp}\Psi_{j}$,  the 
full-dimensional formalism, Eqs.~(\ref{H}) - (\ref{HNL}), can be expressed by 
the following dimensionless coupled equations:
{\small\begin{eqnarray}
\label{eq:gp1d}
\mathrm{i}\frac{\partial \psi_j} {\partial t}&=&-\frac{1}{2}
\frac{\partial^2 \psi_j}{\partial x^2} +
(-1)^j\mathrm{i}\gamma\frac{\partial\psi_j}{\partial x}+\nu_R\;\psi_{3-j} \notag \\ 
&+&\left(\bar g_{jj}\lvert \psi_j \rvert^{2}+\bar g_{12}\lvert\psi_{3-j}\rvert^{2} 
+\bar\chi_{jj} \lvert\psi_j \rvert^{4}\right) \psi_j \\ 
&+& \bar\chi_{12} \left(\lvert\psi_{3-j}\rvert^{2}+2\lvert\psi_{j}\rvert^{2}\right)
\lvert\psi_{3-j} \rvert^{2} \psi_j\;\;(j=1,2)
\notag\end{eqnarray}
}in which the symmetry $\bar g_{12}=\bar g_{21}$, $\bar\chi_{12}=\bar\chi_{21}$ 
is assumed.

\section{Linear stability analysis}
\label{sec:3}
The fundamental framework of MI relies on linear stability analysis, 
such that the steady-state solution is perturbed by a small amplitude/phase, which 
is followed by verifying whether the perturbation amplitude grows or
decays~\cite{Agrawal2013}. {\color{black} Linear stability has been employed in
many different contexts in physics, when looking for relevant effects emerging
from perturbations of existing nonlinear solutions.  
Several phenomena in BEC have been described by considering linear stability analyses.
Among them, we could mention the formation of bright soliton trains in nonlinear 
waves~\cite{2004Carr,2005Abdullaev}, the investigations of Faraday 
waves~\cite{2019-Abdullaev}, and parametric resonances~\cite{2014Cairncross}.
Within our purpose, linear stability is an essential tool to study how MI can
emerge in a nonlinear system.
For that, let us consider that initially the two coupled 
densities of a miscible spin-orbit} BEC are given by
$n_{j}=\lvert\psi _{j0}\rvert^{2}$, with the unperturbed wave function having the 
form $\psi_{j0}=\sqrt{n_{j}}~\mathrm{e}^{-\mathrm{i}\mu t},$ where $\mu$ is the 
chemical potential, common for both components. Then, the stability of 
the coupled system can be examined by assuming that the two wave-function
components of Eq.~\eqref{eq:gp1d} are affected by small space-time perturbations 
$\phi_j\equiv\phi_j(x,t)$, such that they are given by
\begin{align}
\label{perturbation}
\psi_{j}=(\sqrt{n_{j}}+\phi_{j})~\mathrm{e}^{-\mathrm{i}\mu t}.
\end{align}
By using this ansatz, following a procedure as detailed in 
Refs.~\cite{Kasamatsu2004,Kasamatsu2006,2015Bhat}, a simpler relation can be 
derived for the dispersion relation. By assuming that the two pseudo-spin states 
remain with equal densities, $n\equiv n_j$, the perturbations 
$\phi_j\equiv\phi_j(x,t)$ provides all the space-time dependence of the wave functions. 
So, we can further consider the densities within the redefinition of the nonlinear 
interaction parameters, as 
$g\equiv 2n \bar g_{jj}$,
$g_{12}\equiv 2n \bar g_{12}$
$\chi\equiv (2n)^2\bar\chi_{jj}$,
$\chi_{12}\equiv(2n)^2\bar\chi_{12}$. Therefore, with the densities being taken into account within the interaction parameter definitions, the 
coupled equations for the perturbations can be written as 
{\color{black}
{\small
\begin{eqnarray}
{\rm i}\frac{\partial\phi_j}{\partial t}&=&-\frac{1}
{2}\frac{\partial^{2}\phi_j}{\partial x^{2}}+
(-1)^j{\rm i}\gamma\frac{\partial\phi_j}{\partial x}+
\nu_R\left(\phi_{3-j}-\phi_j\right)\notag\\
&+&g\frac{\phi_j +\phi_j^{*}}{2}+ 
g_{12}\frac{\phi_{3-j}+\phi_{3-j}^{*}}{2}\notag \\ 
&+&\chi\frac{\phi_j+\phi_j^{*}}{2}+
\chi_{12}\Big[\frac{\phi_j+\phi_j^{*}}{2}
+\left(\phi_{3-j}+\phi_{3-j}^{*}\right)\Big], \nonumber\\
{\rm i}\frac{\partial\phi_j}{\partial t}&=&-\frac{1}{2}\frac{\partial^{2}\phi_j}{\partial x^{2}}+
(-1)^j{\rm i}\gamma\frac{\partial\phi_j}{\partial x}+
\nu_R\left(\phi_{3-j}-\phi_j\right)\label{A20}\\
&+&(g+\chi+\chi_{12})\frac{\phi_j +\phi_j^{*}}{2}+ (g_{12}+2\chi_{12})\frac{\phi_{3-j}+\phi_{3-j}^{*}}{2}\notag .
\end{eqnarray}} 
In view of the above final structure of the coupled equations,
we also notice the convenience of redefining the
nonlinear parameters as 
$\alpha\equiv g+\chi+\chi_{12}$ and  
$\beta\equiv g_{12}+2 \chi_{12}$. So,  with 
$\Re(\phi_j)$ being the real part of $\phi_j$, we can rewrite \eqref{A20} as
}

{\small
\begin{equation}
{\rm i}\frac{\partial\phi_j}{\partial t}=-\frac{1}{2}\frac{\partial^{2}\phi_j}
{\partial x^{2}}+
(-1)^j{\rm i}\gamma\frac{\partial\phi_j}{\partial x}+\nu_R \left(\phi_{3-j}-\phi_j\right)
+ \Re\left(\alpha\phi_j+\beta\phi_{3-j}\right),\label{A2}
\end{equation}
}where, for any complex $f$,
$\Re(f)$ represents the real part of $f$. 
For the space-time dependence of $\phi_j$ we assume the general complex form 
\begin{align}\label{A3}
\phi_{j}=\zeta_{j} \cos\left(k x-\Omega t\right)+\mathrm{i} \eta _{j} 
\sin\left(k x-\Omega t\right),
\end{align}
where $\zeta _{j}$ and $\eta _{j}$ are, respectively, {\color{black}
the amplitudes for the real and imaginary parts of $\phi_j$, which have spatial-time oscillations 
given by the wave number $k$ and eigenfrequency $\Omega$. In correspondence to our space and time units, 
$k$ and $\Omega$ are in units of $1/\ell_\perp$ and $\omega_\perp$, respectively.}

\subsection{Dispersion and modulational instability relations}
The MI relations, with their dependence on the linear and nonlinear parameters, 
are derived by solving a matrix equation, once considered the general form 
\eqref{A3} in the coupled Eq.~\eqref{A2}. 
{\color{black}
In our analysis, we start by considering 
parametrizations such that the eigenfrequency solutions have no imaginary parts. 
Within the changes in the parameters, MIs can emerge represented by the imaginary
parts of the $\Omega$ solutions. 
Within an experiment realization, this is clearly explained in the first paragraphs 
of Ref.~\cite{2017Nguyen}. The rapid growth of the fluctuations 
can lead to a breakup in wave propagation.}

Therefore, by assuming the perturbation given by Eqs.~\eqref{A2} and~\eqref{A3}, two 
inter-dependent equations for the real and imaginary parts of $\phi_j$ are
obtained, as follows:
{\footnotesize
\begin{eqnarray}
\hspace{-.5cm}\left[
\begin{array}{cc}\Omega-k\gamma&0\\0&\Omega+k\gamma\end{array}
\right]
\left[
\begin{array}{c}\zeta_1\\\zeta_2\end{array}
\right]
&=&
\left[
\begin{array}{cc}\frac{k^2}{2}-\nu_R &\nu_R \\\nu_R &\frac{k^2}{2}-\nu_R \\\end{array}
\right]
\left[
\begin{array}{c}\eta_1\\\eta_2\end{array}
\right]\label{A4}\\
\hspace{-.5cm}\left[
\begin{array}{cc}\Omega-k\gamma&0\\0&\Omega+k\gamma\end{array}
\right]
\left[
\begin{array}{c}\eta_1\\\eta_2\end{array}
\right]
&=&
\left[
\begin{array}{cc}\frac{k^2}{2}-\nu_R +\alpha&\nu_R +\beta\\
\nu_R +\beta&\frac{k^2}{2}-\nu_R +\alpha\\\end{array}
\right]
\left[
\begin{array}{c}
\zeta_1\\
\zeta_2
\end{array}
\right]\nonumber .
\end{eqnarray}
}By solving the above-coupled matrix equation, four solutions emerge
for $\Omega$, functions of $k^2$, given by $\pm\Omega_\pm(k^2)$.
Parameterized by the SO ($\gamma$) and Rabi coupling ($\nu_R$), they are given by 
{\small\begin{eqnarray}
\Omega_\pm^2(k^2,\nu_R,\alpha,\gamma)\equiv
{\cal F}(k^2,\nu_R,\alpha,\gamma)\pm {\cal G}(k^2,\nu_R,\alpha,\gamma),\label{Omega2}
\end{eqnarray}}
where
{\footnotesize\begin{eqnarray}
{\cal F}(k^2,\nu_R,\alpha,\gamma)&\equiv& (k\gamma)^2+\nu_R (\nu_R +\beta)+ 
{\cal K}(k^2,\nu_R,\alpha)
\nonumber\\
\left[{\cal G}(k^2,\nu_R,\alpha,\gamma)\right]^2
&\equiv&\left[
\left(2\nu_R +{\beta}\right)\left(\frac{k^2}{2}-\nu_R\right)+\nu_R\alpha\right]^2
\label{FGK}\\
&+&4(k\gamma)^2{\cal K}(k^2,\nu_R,\alpha),\nonumber\\
{\cal K}(k^2,\nu_R,\alpha)&\equiv&\left(\frac{k^2}{2}-\nu_R \right)
\left(\frac{k^2}{2}-\nu_R +\alpha\right).\nonumber
\end{eqnarray}}
These solutions for $\Omega$ can be real or complex, with the signs of the real and  
imaginary parts (positive or negative) depending on the interaction parameters, 
which are reflecting the miscibility of the mixture, as well as the strengths of the 
SO and Rabi couplings. 
The explicit dependence on the parameters shown in \eqref{Omega2}-\eqref{FGK} is
avoided in the next.

{\color{black} 
With the instability growth rates given by the respective imaginary part solutions of $\pm\Omega_\pm$, 
in the specific experimental study described in \cite{2017Nguyen}, the formation of soliton train solutions 
arises with the wave breakup due to the MI rapid growth of fluctuations}. 
In view of the symmetry of our assumption for the perturbations \eqref{A3}, only two are 
relevant in our analysis, implying that $\Omega$ is real positive in the 
initial assumption \eqref{A3}. From this equation, one can also anticipate the 
symmetry found in the results with respect to the wavenumber $k$. Therefore, as we 
are concerned only with the possible instabilities which can occur, our analysis 
will be concentrated on the cases in which we have non-zero imaginary parts 
of $\Omega_\pm$. Given the ${\cal F}$ and ${\cal G}$ defined 
in Eqs.\eqref{Omega2}-\eqref{FGK}, and considering the symmetry of the 
possible solutions, the growth of 
instabilities (gain), $\xi_{\pm}\equiv\xi_{\pm}(k^2)$, can be expressed by    
{\footnotesize\begin{eqnarray}
&&\xi_{\pm}=
\left|{\rm Im}\left({\Omega_\pm}\right)\right|=
\left|{\rm Im}\left(\sqrt{{\cal F}\pm {\cal G}}\right)\right|>0,\label{imag}
\end{eqnarray}}with the coupled system being stable if $\xi_{\pm}=0$.
{\color{black} Here, we should stress that for absolute stable solutions, both $\xi_{\pm}$ 
must be zero. In all the solutions that we are reporting, the stability was confirmed by 
direct calculation of the corresponding GP equation.}
By inspecting the above expression for $\Omega_\pm$, we noticed
that $\xi_{\pm}(k^2)$ are symmetric with respect to the sign of the wave number $k$, 
being also independent of the sign of $\gamma$ (which is related to the laser momentum direction in the 
coupling of the two components). 

\subsection{On the miscibility with cubic and quintic interactions}

Before considering the treatment of a coupled mixture of two condensed systems,
let us mention the simpler case of single-component, which is the case
that a two-component mixture (such as two hyperfine states of the same atom) is 
reduced in the absence of linear couplings (SO and Rabi), when the inter-species 
interactions are set to zero ($g_{12}=0$ and $\chi_{12}=0$). This single-component
case was already covered in Ref.~\cite{Kasamatsu2006}, when considering binary 
attractive interactions. In our case with quintic nonlinear interactions, 
this single-component condition for non-zero MI is realized when
{\small\begin{eqnarray}
\xi\Big|_{sc}&=&{\rm Im}\sqrt{\frac{k^2}{2}\left(\frac{k^2}{2}+g+\chi\right)}\ne 0,
\label{sc}
\end{eqnarray}
}implying that the previous study in Ref.~\cite{Kasamatsu2006} is affected only by 
a rescaling in the two-body interaction parameter, when considering a quintic term. 
For example, there is no $k^2-$region with MI when we have only repulsive two-body
interactions ($g>0$). However, in the same $k^2-$region MI can be introduced with 
a sufficiently larger attractive three-body interaction, such that $|\chi|>g$.
As an extension, the possible MI generated by attractive two-body ($g<0$) 
interaction can be reduced (or removed) with a repulsive three-body interaction 
such that $\chi>|g|$. 
In addition, we can anticipate that our results are also in perfect agreement 
with some other previously known MI studies for single-component models 
provided in Refs.~\cite{Agrawal2013,Theocharis}. 

For a coupled two-component system, we can mention a previous study 
in Ref.~\cite{2015Bhat}, obtained in the absence of three-body interactions 
($\chi=0$ and $\chi_{12}=0$), in which the authors reported results that
can be reproduced by our extended study. In practice,
by going beyond the work done in Ref.~\cite{2015Bhat}, with the addition of 
quintic interactions, we are also investigating the interplay between cubic, 
quintic and linear coupling terms, with a particular focus on miscibility 
conditions of the mixture.

From the expression for the eigen-frequency solutions $\Omega^2$ \eqref{Omega2}, 
the relevance of the miscibility can become more apparent with an appropriate
redefinition of the parameters $\alpha$ and $\beta$, such that the intra- and
inter-component interactions could be well distinguished, by noticing that
$\alpha -\chi_{12}=g+\chi$ and $\beta -\chi_{12}=g_{12}+\chi_{12}$.
Therefore, when considering cubic and quintic nonlinear parameters, 
the relative dominance between inter- and intra-component interactions,
can be measured by their difference. Within our assumption, 
it can be  defined by
\begin{equation}
\Lambda\equiv \beta-\alpha = ({g}_{12}+\chi_{12})-({g}+\chi),\label{Lambda}
\end{equation}
This parameter $\Lambda$ provides an extension of the miscibility parameter previously 
derived for homogeneous mixtures of quantum fluids when assuming 
only cubic nonlinear interactions, expressed by $\Delta\equiv 
\bar g_{12}^2/(\bar g_{11}\bar g_{22})$~\cite{2008Pethick}
(also considered in Ref.~\cite{2005Merhasin}, with no linear couplings).
As indicated by $\Delta$, in which an immiscible configuration is given by $\Delta>1$,
the corresponding conditions, when assuming two- and three-body interactions, 
$\Lambda>0$ indicates that the coupled system becomes more immiscible; 
with $\Lambda<0$ pointing out a more miscible configuration.
However, such simplified criteria for the miscibility transition ($\Delta$ or $\Lambda$),
valid for homogeneous coupled systems, can be affected when considering other 
effects in the mixture, as the linear SO and Rabi couplings.
So, a more reliable quantitative criterion for the miscibility can be directly obtained 
by the corresponding density distributions of the two species, with their possible 
overlap being affected by the non-homogeneous characteristics of the system.
For this purpose, in Refs.~\cite{Wen2012,Kishor2017,Gutierrez2021} different 
definitions for miscibility have been proposed based on the spatial overlap 
between the two atomic clouds. However, these definitions can be more useful
in the final analysis of MI results as related to the miscibility. For such 
purpose, we can assume the miscibility factor is based on the absolute values 
of the wave functions, given by
$\eta=\int dx |\psi_1||\psi_2|,$
which is 1 in the case of complete overlap between the two densities; 
and zero when they are completely separated.
The instability of the SO coupled-BECs for different system parameters 
can be scrutinized in both cases of miscible and immiscible regimes. 

In the following section, we provide an analysis of the MI, with corresponding results, 
when considering cubic and quintic interactions. Mostly, the results regarding to the 
nonlinear interactions are expressed in terms of the parameters $\alpha$ and $\beta$, which 
are found convenient when considering both nonlinear two- and three-body interactions, 
as their difference is directly related to the miscibility of the mixture, 
before considering the linear couplings. 

\section{Results on the MI with cubic-quintic interactions
and linear couplings}
\label{sec:4}
With the expressions \eqref{Omega2}-\eqref{imag} providing the MI gain, we present 
in this section an analysis of different possibilities, when considering the 
linear couplings with the two- and three-body interactions, together with some 
sample significant results. 
Among the simplest limiting cases that we may consider, we notice that when 
the nonlinearities are such that $\alpha=\beta=0$ (implying a balance between 
repulsive and attractive nonlinear interactions, or when they are null), the 
solutions are real, expressed by the linear dispersion,
\begin{eqnarray}
\Omega_\pm=\frac{k^2}{2}-\nu_R \pm\sqrt{\nu_R ^2+(k\gamma)^2},
\label{linear}\end{eqnarray}
with no instabilities ($\xi_\pm=0$).
The interesting cases, in which the nonlinearities, together with the linear
couplings, can produce relevant effects due to modulational instabilities 
is being discussed in the following. First, we are selecting four particular
limiting cases, with corresponding diagrammatic representations. In the final
part of the section, a more general picture is presented.\\

\noindent {\bf (i) Zero wave-number limit ($k=0$).}\\
Within all the solutions, the specific $k=0$ limit, 
for which $\Omega^2=2\nu_R (2\nu_R +\beta-\alpha)$, is of interest to verify 
the possible nonzero MI solutions in the extreme limit where the perturbations 
are close to zero kinetic energy. In this limiting case, one cannot expect any effect 
coming from the SOC, considering that it is associated with the first derivative of $x$. 
So, MI can happen only for one of the cases given by \eqref{imag}, 
{\small\begin{equation}
\begin{array}{l} 
\xi\Big|_{k=0}=\left|{\rm Im}\left(\sqrt{2\nu_R (2\nu_R +{\beta-\alpha})}\right)\right|
\end{array},\label{k0}
\end{equation}
}when the system is more miscible, with $\alpha-\beta$ enough larger than the Rabi 
frequency parameter, which is assumed positively defined.
There is no loss of generality in assuming 
$\nu_R$ always positive, as we can shortly explain:

Let us consider the two possibilities for $\nu_R$:  
if $\nu_R>0$, MI can happen only for $(\beta-\alpha)<-2\nu_R $; whereas, if 
$\nu_R<0$, it could happen for $(\beta-\alpha)>-2\nu_R $. 
However, this apparent problem just 
reflects the fact that the parametric regions in which the MI results are located
are changing the positions according to our choice for the direction of the Rabi 
coupling. Once given the two and three-body interactions, the unstable and
stable regions are visualized in different positions of the parametric space
defined by these interactions. 
\begin{figure}[!ht]
\begin{center}
\includegraphics[width=4.3cm,height=3.9cm]{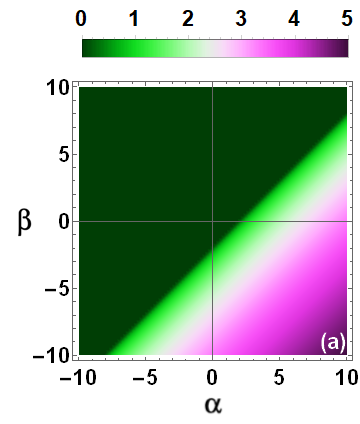}
\includegraphics[width=4.cm,height=3.7cm]{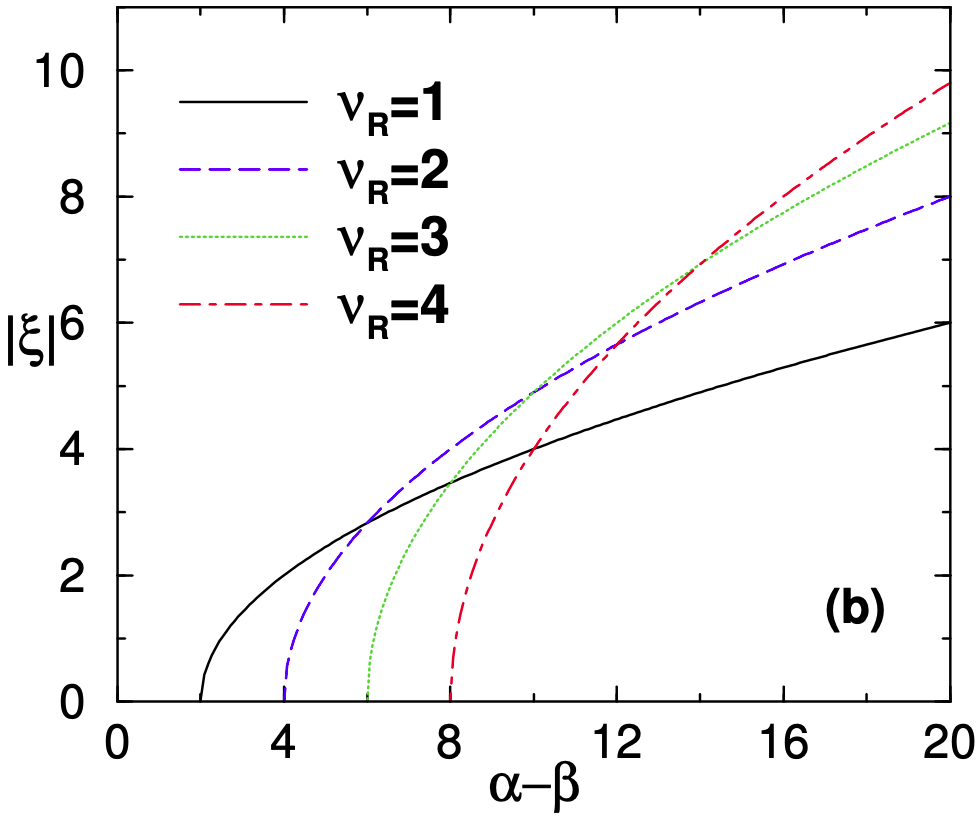}
\end{center}
\vspace{-.5cm}
\caption{(Color online) In panel (a), for $k=0$ and $\nu_R=1$, it is shown the only possible 
parametric region of MI (with the $|\xi|$ strengths indicated by the color bar) 
defined by the interactions $\alpha$ and $\beta$ (determined by  Eq.~\eqref{k0}). 
Panel (b) shows how $|\xi|$ evolves with $\alpha-\beta=(g+\chi)-(g_{12}+\chi_{12})$, 
for different values of $\nu_R$. All quantities are dimensionless.}
\label{f01}
\end{figure}

The present case with $k=0$ is illustrated by two panels in Fig.~\ref{f01}.
Panel (a) provides the parametric region defined by $\alpha$ and $\beta$, considering
$\nu_R=1$, with panel (b) showing how the MI ($|\xi|>0$) increases with the
miscibility, considering the parameter $\Lambda=\beta-\alpha$ going to more negative values, 
with different strengths of the Rabi oscillations.
The MI can only occur when the coupled system is in a more miscible configuration, given by 
$\Lambda<-2\nu_R$: With the inter-species interactions weaker than the intra-species ones,
if we assume all the nonlinear interactions are repulsive; or, with the inter-species more 
attractive than the intra-species, in the case of net-attractive interactions.
This $k=0$ limit can be verified as a particular point within the next cases that we are 
going to discuss.
Note that the above result for two components, implying the existence of MI when $k=0$, 
is only relying on small linear Rabi coupling. By increasing $\nu_R$ (strong Rabi coupling) 
the parametric region for MI is reduced, as shown in Fig.~\ref{f01}. In the limit of
zero $\nu_R$ the result matches with the 
single-component case, shown in \eqref{sc}, where MI can occur only for $g+\chi<-k^2/2$.\\

\begin{figure}[!ht]
\centering\includegraphics[width=4.3cm,height=4cm]{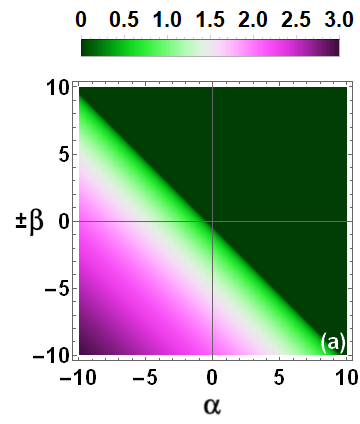}
\centering\includegraphics[width=4.1cm,height=3.7cm]{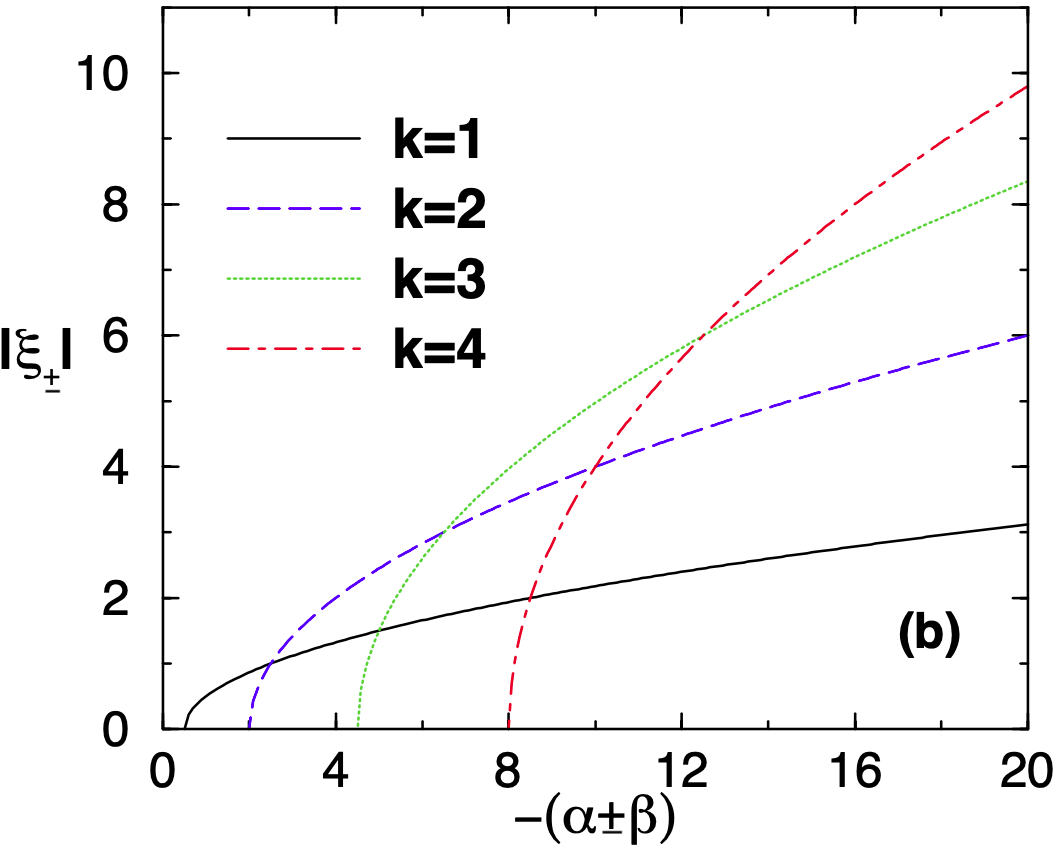}
\caption{(Color online) Without SO and Rabi couplings ($\gamma=0$, $\nu_R=0$), 
as given by Eq.~\eqref{xi00}. In panel (a) it is shown the MI parametric region for 
$|\xi_+|$ (given by $+\beta$ vs $\alpha$) and $|\xi_-|$(given by $-\beta$ vs $\alpha$), 
with $k=1$ (the $|\xi_\pm|$ strengths are indicated in the color bar). 
The panel (b) shows the corresponding behavior of $\xi_{\pm}$, in terms of
$-(\alpha\pm\beta)$, for a few values of $k$, as indicated.
All quantities are dimensionless.}
\label{f02}
\end{figure}

\noindent {\bf (ii) \bf Zero linear couplings ($\gamma=0$ and $\nu_R =0$).}\\
With non-zero nonlinearity, the simplest case of interest to be analyzed, 
in view of possible MI occurrence, is when there is no linear coupling, such that the SO and 
Rabi couplings are zero ($\gamma=0$ and $\nu_R =0$, respectively). 
In this particular case, the coupling is given by the inter-species interaction, with the 
possible non-zero MI solutions happening only when $(\alpha\pm\beta)<0$, with 
$2|\alpha\pm\beta|>k^2$, with the growth of instabilities given by
\begin{eqnarray}
\xi_{\pm}\Big|_{(0,0)}&=&\left|{\rm Im}\left(
\frac{k}{2}\sqrt{{k^2}+2(\alpha\pm\beta)}
\right)\right|.\label{xi00}
\end{eqnarray}
The two solutions are affected by the relative values of the 
nonlinear parameters $\alpha$ and $\beta$, with 
$\alpha+\beta= g+ g_{12} +\chi+3\chi_{12}$
and $\alpha-\beta= g+\chi- g_{12}-\chi_{12}$.
So, we can have $\xi_\pm\ne 0$ for $(\alpha\pm\beta)<0$ with $|\alpha\pm\beta|>k^2/2$. 
The corresponding parametric regions given by $\alpha$ and $\beta$ are shown in 
panel (a) of Fig.~\ref{f02} for $k=1$. The increasing behavior of the instabilities with 
$k$ is also pointed out in panel (b) of this figure.

\noindent {\bf (iii) Non-zero Rabi with zero SO couplings ($\nu_R\ne 0, \gamma=0$).}\\
When the linear coupling is only via Rabi frequency, without SOC, 
($\gamma=0$, with $\nu_R \ne 0$), the possible MI  
non-zero solutions are given by
{\footnotesize\begin{eqnarray}
\xi_{\pm}\Big|_{(0,\nu_R )}\hspace{-0.3cm}=\left\{
\begin{array}{l}
\left|{\rm Im}\left(
\frac{k}{2}\sqrt{{k^2}+2(\alpha+\beta)}
\right)
\right|
\\ \\
\left|{\rm Im}\left(
\sqrt{
\left(\frac{k^2}{2}-2\nu_R \right)
\left[\frac{k^2}{2}-2\nu_R +\alpha-\beta\right]
}\right)
\right|\end{array}\right.
,\label{xi01}
\end{eqnarray}
}where we noticed that $\xi_+|_{0,\nu_R}$ is not affected by $\nu_R$, 
such that the result is identical to $\xi_+|_{0,0}$
given in \eqref{xi00}.
\begin{figure}[!h]
\includegraphics[width=4.3cm,height=4cm]{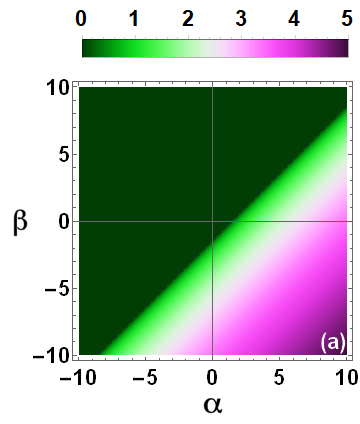}
\includegraphics[width=4.1cm,height=3.7cm]{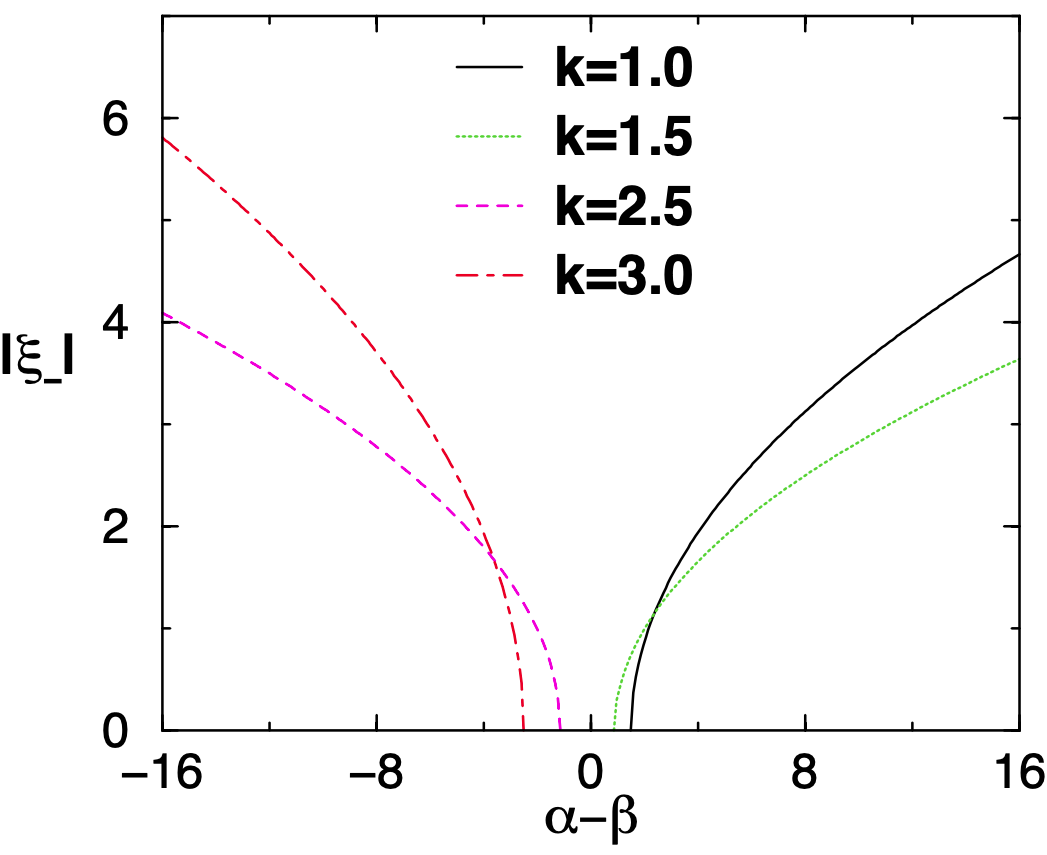}
\caption{(Color online) 
In panel (a), for $k=1$, the MI region for $|\xi_-|$ is shown for the case 
without SO ($\gamma=0$) and with Rabi ($\nu_R=1$) couplings, parameterized by 
$\alpha$ and $\beta$ (with the strength given by the color bar), as given by 
\eqref{xi01}. 
The results for $|\xi_+|$ (not depending on $\nu_R$) are already in 
Fig.~\ref{f02}. Panel (b) shows how $|\xi_-|$ 
evolves with $k$, as indicated ($k=1$ corresponding to the left panel). 
As shown, the MI can happen for both signs of $\alpha-\beta$. 
All quantities are dimensionless.}
\label{f03}
\end{figure}

The Rabi constant is affecting only $\xi_-$,
with $\xi_+$ not depending on the Rabi constant being identical to
\eqref{xi00}.
$\xi_+$ can be non-zero for $\alpha+\beta<0$ with  
$2|\alpha+\beta|>k^2$.  The parametric region, when considering $\nu_R=1$ and $k=1$, 
is shown in Fig.~\ref{f03}. Also, one should notice that MI can occur for stronger 
Rabi coupling ($\nu_R\gg k^2$), by considering $\xi_-$ in \eqref{xi01} with $\alpha-\beta>2\nu_R$.

\begin{figure}[!h]
\centering\includegraphics[width=0.95\linewidth]{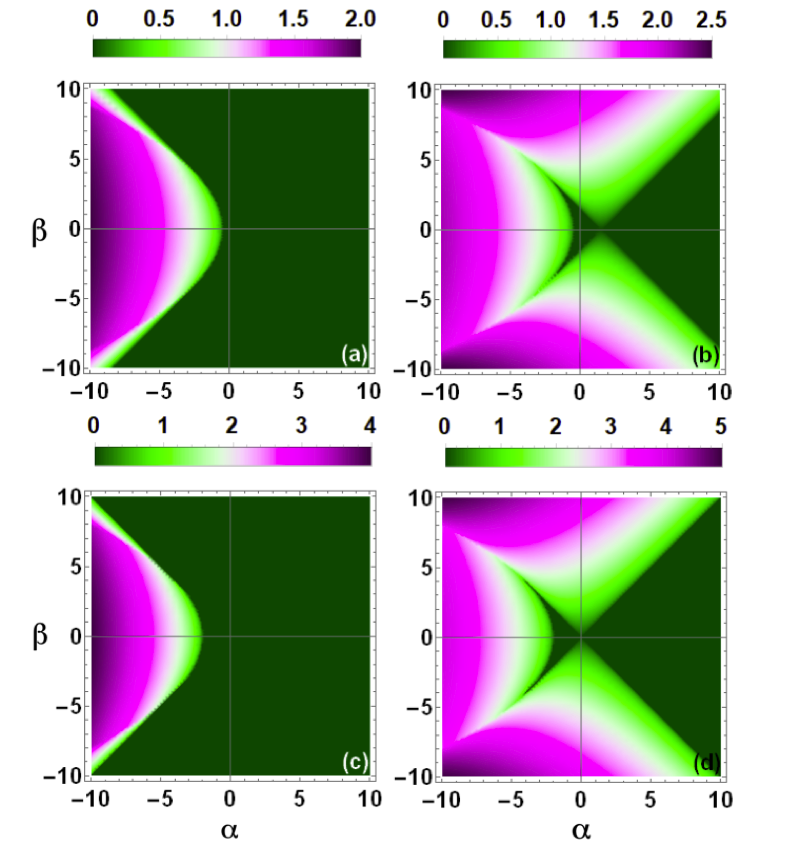}
\caption{(Color online)
The MI $\xi_\pm$ regions parameterized by $\alpha$ and $\beta$, as given  
by~\eqref{caseD}, are shown for the case with SO ($\gamma=1$) and 
without Rabi ($\nu_R=0$) couplings.
The left (right) panels are for $\xi_{+}$ ($\xi_{-}$), with  
the upper (lower) panels for $k=1$ ($k=2$).
The strengths of $|\xi_\pm|$ are, respectively, indicated by the color bars. 
All quantities are dimensionless.}
\label{f04}
\end{figure}
\noindent {\bf (iv) Non-zero SO with zero Rabi couplings ($\gamma\ne 0, \nu_R  =0$).}\\
When the linear coupling occurs only via SO, from \eqref{Omega2} the MI expression
is given by 
{\footnotesize \begin{eqnarray} \color{black}
\xi_{\pm}\Big|_{(\gamma,0)}=
\left|{\rm Im}\sqrt{
{\frac{k^2}{2}\left[
\frac{k^2}{2}+\alpha+2\gamma^2\pm \sqrt{
{\beta^2}+8\gamma^2\left({\frac{k^2}{2}+\alpha}\right)}\right]}}
\right| \label{caseD}
,\end{eqnarray}
}with the corresponding parametric regions being illustrated in Fig.~\ref{f04} 
for $\xi_+$ (left panels) and $\xi_-$ (right panels), in which the SO are
assumed fixed to $\gamma=1$. 
In order to verify how the MI regions are affected 
by changing the wave numbers, we consider $k=1$ in the panels (a) and (b), with $k=2$ 
in the panels (c) and (d).
As seen, the regions defined by $\alpha$ and $\beta$ where MI occurs become slightly 
more complex than the previously analyzed ones, particularly for the case of $\xi_-$,
as the region separations are not always linear. In the case of $\xi_+$, the 
instabilities appear only for $\alpha<0$, in a symmetric way for $\beta$ positive
or negative. 
However, for $\xi_-$ the stable regions are quite reduced, with the superposition 
of three branches with modulational instabilities, as seen in the right panel of
Fig.~\ref{f04}.
Therefore, even with $\alpha+2\gamma^2>0$, for the particular case of $\xi_-$,
we can have imaginary terms arising from the argument of the internal 
square-root term, for enough larger values of $\beta$.
These sample results given in Fig.~\ref{f04} are already indicating that we 
should expect an increasing complexity in the parametric pictures defining 
the regions with MI.
However, we still noticed that the effect of the three-body quintic term in the 
definition of the parametric regions can be hidden in a 
redefinition applied to the previously adjusted two-body parameters, by keeping the
same values for $\alpha$ and $\beta$. These two parameters, by combining cubic and
quintic nonlinear interactions with respective densities are essentially 
representing the proportional effects emerging from two- and three-body in the 
modulational instabilities.

In the following, by analyzing more general results, 
our aim is to clarify possible relevant aspects related to the
role of three-body quintic terms in the MI when considering SO and Rabi couplings, 
as well to point out  interesting aspects of the interplay between the linear 
couplings and interactions that would be explored in experimental realizations.\\

\noindent {\bf (v) More general cases with both linear couplings and cubic-quintic interactions.}\\
The previous four cases focusing on the parametric regions of MI are exemplifying 
particular conditions in which some of the coupling parameters are set to zero,
in order to estimate their role in more general situations in which both 
couplings are participating. We follow this section by considering the interplay between 
cubic and quintic interactions with the linear SO and Rabi couplings. 

\begin{figure}[!h]
\centering\includegraphics[width=0.95\linewidth]{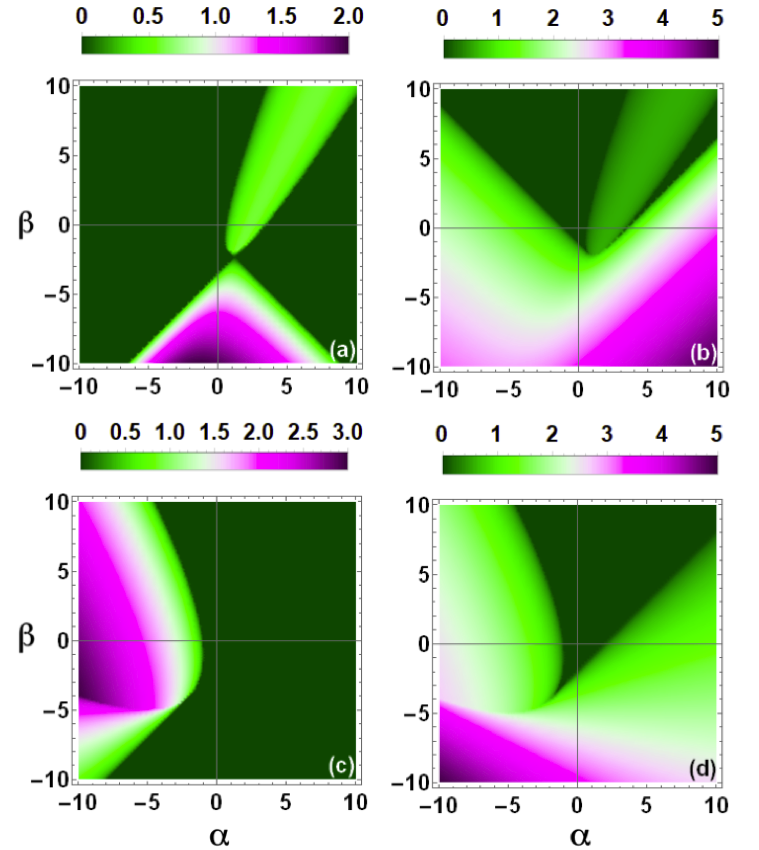}
\caption{(Color online) 
For fixed SO ($\gamma=1$) and Rabi ($\nu_R =1$) couplings,
we have the MI $\xi_\pm$ regions (with color bars indicating their
magnitudes), parameterized by the interactions $\alpha$ and 
$\beta$ (through Eq.~\eqref{imag}),
The left (right) panels are for $\xi_{+}$ ($\xi_{-}$), with  
the upper (lower) panels being for $k=1$ ($k=2$). 
All quantities are dimensionless.}
\label{f05}
\end{figure}

\begin{figure}[!ht]
\centering\includegraphics[width=0.99\linewidth]{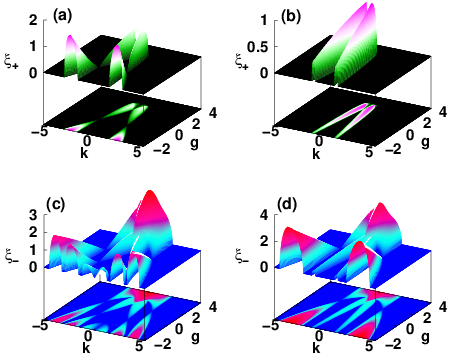}
\caption{(Color online) Surface 3D plots  
showing the MI gain $\xi_\pm={\rm Im}(\Omega_\pm)$ [(a) and (b) for 
$\xi_{+}$, with (c) and (d) for $\xi_{-}$], obtained by solving Eq.~(\ref{imag}), 
{\color{black} They are given as functions of the wave-number $k$ and 
two-body intra-species interaction $g\equiv g_{jj}$, for the inter-species fixed at $g_{12}=1$. 
The SOC and Rabi parameters are fixed, respectively, to $\nu_R =1$ and $\gamma=1$.
With $\chi_{ij}= 0$, in panels (a) and (c); and, respectively, $\chi_{ij}=1$, in panels (b) and (d),
one can appreciate the effect of quintic (three-body) nonlinearity in the MI gain.}
All the quantities are dimensionless.}
\label{f06}
\end{figure}
\begin{figure*}[!ht]
\centering\includegraphics[width=0.9\linewidth]{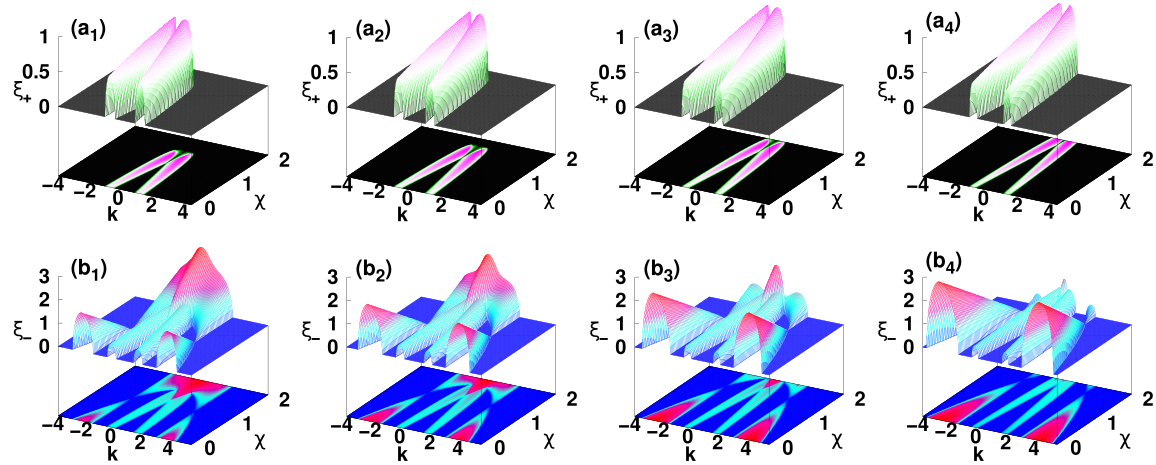}
\caption{(Color online) Surface 3D plots (with corresponding 2D projections) are 
shown for the MI gain $\xi_+$ (top row) and $\xi_{-}$ (bottom row) in planes defined
by repulsive three-body intra-species interactions $\chi$ and the wave number $k$.
In each column, the interspecies two-body interaction $g_{12}$ is fixed,  such that
$g_{12}=0.1$ [panels 
(a$_1$) and (b$_1$)], $g_{12}=0.5$ [panels (a$_2$) and (b$_2$)],
$g_{12}=1.0$ [panels (a$_3$) and (b$_3$)], and $g_{12}=1.5$ [panels (a$_4$) and (b$_4$)].
The remaining parameters are fixed, with $g=1$, $\chi_{12}=1$, $\nu_R =1$, and $\gamma=1$.
All quantities are dimensionless.}
\label{f07}
\end{figure*}

As to compare with the particular cases discussed in the previous section, in
Fig.~\ref{f05} we present the parametric regions for MI, defined by the interactions
$\alpha$ and $\beta$, in which both can be repulsive or attractive. Besides being fixed 
the linear couplings, with $\nu_R=1$ and $\gamma=1$, this can already give  
a picture of more general situations in which both SO and Rabi couplings are 
being considered, with their variations governed by Eqs.~\eqref{Omega2}-\eqref{imag}.
Therefore, we are providing two panels in Fig.~\ref{f05}, respectively 
for $\xi_+$ and $\xi_-$ as defined by Eqs.~\eqref{Omega2}-\eqref{imag}.
{\color{black}
The diagrams should be exactly reproduced if just cubic nonlinear terms have been 
taken into account, with the replacement $\alpha\to g$ and $\beta\to g_{12}$.}
These considerations led us to the conclusion that, to a certain extent, the 
inclusion of quintic nonlinear terms in the GP formalism implies a renormalization 
of the parameters, such that three-body effects could be considered effective inside 
the previous two-body nonlinear parameters. 
However, this is not generally true, as the effect of quintic interactions in the 
theory can also change significantly the results previously verified for the MI 
gain; particularly, when such effects come with different signs (attractive or 
repulsive), as we can verify in a few examples that will be presented. Even when 
a parameter rescaling can lead to the same result, it will be of interest to 
distinguish the separate contributions provided by cubic or quintic interactions.
Our following results and analysis are in this regard.

\begin{figure}[!ht]
\centering\includegraphics[width=1.0\linewidth]{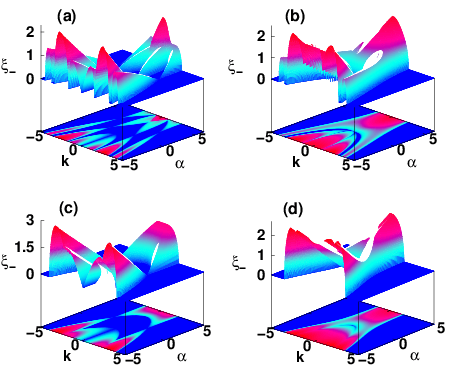}
\caption{(Color online) 
Rabi and SO coupling effects are shown with four different combinations,
by surface 3D plots (with corresponding 2D projections) for the MI gain $\xi_-$,
in planes defined by $k$ versus $\alpha$, with fixed $\beta=1$.
($\nu_R $, $\gamma$)= (1,1) [panel (a)],
(0,1) [panel (b)],  
(1,0) [panel (c)], and
(0,0) [panel (d)]. 
All quantities are dimensionless.}
\label{f08}
\end{figure}

In Fig.~\ref{f06}, through four panels, we show the $\xi_\pm$ MI gains as functions of $k$ 
and the intra-species parameter $g$, in which we consider fixed values for the SO and 
Rabi couplings ($\gamma=1$, $\nu_R =1$), as well as for the inter-species parameter 
($g_{12}=1$). In panels (a) and (c), all three-body parameters are neglected $ \chi_{ij}=0$; 
whereas in panels (b) and (d), they are non-zero with $ \chi_{ij}=1$. 
These sample comparison results are just to see the kind of effect we can have, 
due to the modulational instabilities, which can arise when taking into account
nonlinear three-body effects in previously obtained results, as the ones of  
Ref.~\cite{2015Bhat}. In this case, the quintic effect introduced is 
The results presented in panels (a) and (c), where $\chi=\chi_{12}=0$ and $\nu_R=\gamma=1$ 
(only cubic terms are considered for the nonlinear interactions), refer to $\alpha= g$ 
and $\beta= g_{12} =1$. With these parameters, $\xi_+\ne 0$ and we 
have solutions only for a small window given by $\nu_R >\frac{k^2}{4}$, as shown 
in Eq.~\eqref{xi01}.
{\color{black}
Here, becomes clear that we could manipulate the parameters such that the same plotted results 
could be obtained with non-zero three-body parameters, as we just need to keep the same 
$\alpha$ and $\beta$.} Suppose, for example, that $g_{12}=0.5$, such that $\chi_{12}=0.25$, 
$\beta$ remains equal to 1. In the same way, we fix $\alpha=g+\chi+\chi_{12}=2$, with 
$g+\chi=1.75$.

With continuous variations of the intra-species quintic term interaction ($\chi$), 
as functions of $k$, by increasing the strengths of the inter-species cubic term, 
from $g_{12}=0.1$ till $g_{12}=1.5$, we have the four sets of panels in 
Fig.~\ref{f07} for $\xi_+$ [(a$_1$) to (a$_4)$] and $\xi_-$ [(b$_1$) to (b$_4)$]. 
Reflecting the complexity of the corresponding equations that define $\xi_+$ and 
$\xi_-$, more unstable regions occur for $\xi_-$, with the MI essentially 
growing according to the differences between the three-body $\chi$ 
and two-body $g_{12}$ (both, repulsive interactions). By decreasing 
$\chi$ from 2 to zero, four regions with MI are manifested symmetrically from 
$-k$ to $+k$, with stability at $k=0$. However, on the other side, 
for larger $\chi$ and smaller $g_{12}$, the instability grows mainly 
concentrated close to $k=0$.

In the four panels of Fig.~\ref{f08}, by assuming $\alpha$ from negative
(attractive) to positive (repulsive) interactions, with a fixed repulsive
$\beta=1$, we are verifying the effect of miscibility together with the linear 
couplings on the MI provided by $\xi_-$. 
By considering each panel, varying $\alpha$ from -5 to 5, the miscibility increases
from immiscible configurations [$\Lambda>0$ for $\alpha<1$] to miscible configurations
[$\Lambda<0$ for $\alpha>1$]. Considering more miscible configurations,
the MI turns out to be more concentrated near small values of $k^2$; whereas, in case
of immiscible configurations ($\alpha<1$) several valleys of stability emerge separating 
the MI regions, with the number of MI regions following the complexity of the
$k^2$ polynomial behavior expressions.
The four-panel results shown in Fig.~\ref{f08} are also relevant to get 
a general picture of the role of the linear Rabi and SO couplings, which are
respectively represented by $(\nu_R,\gamma)=$(1,1) [panel (a)],
$(\nu_R,\gamma)=$(0,1) [panel (b)], $(\nu_R,\gamma)=$(1,0) [panel (c)], and
$(\nu_R,\gamma)=$(0,0) [panel (d)]
As noticed, by going from panel (a) to panel (d), the strengths of $\xi_-$ increase 
as the couplings are decreasing.
However, the number of maxima that emerges as $k$ varying increases with the 
linear couplings, as we can verify in the planes defined by $\alpha$ and $k$,
determined by the MI expressions as functions of $k$.

\section{Conclusions} 
\label{sec:conclusion}
Modulation instability modes are studied in binary coupled systems with 
engineered spin-orbit and Rabi-coupled Bose-Einstein condensates, by considering
two- and three-body nonlinear interactions in the Gross-Pitaevskii formalism. 
Our investigation provides a clear picture of the interplay between linear
and nonlinear couplings for the modulation instabilities, considering miscible 
and immiscible configurations, highlighting the specific role of the nonlinear 
cubic-quintic contributions, which goes beyond previous analyses that were 
limited to particular cases in which only some aspects were considered.
For that, we are emphasizing the relevance of the miscibility aspects of the mixture
in generating the instabilities. 

By considering wide variations of the inter- and intra-species interactions, 
going from negative to positive values, 
the parametrizations shown in the present work are expected to be useful in 
experimental studies considering hyperfine levels of the same atomic species, 
whenever the interests are to investigate relevant effects derived from 
modulation instabilities in atomic mixtures.
In several cases, one can verify that the inclusion of a quintic nonlinear term
can be translated by a rescaling of previously considered cubic nonlinear terms, as 
concerned with the overall effect. However, it is relevant to estimate in real experiments
the effect of three-body interactions in addition to the already known two-body ones, in 
the perspective that both interactions could be controlled in an independent way, as pointed out in 
Ref.~\cite{2021Hammond}. In this regard, MI can be studied in a more general context, 
looking for appropriate parametrizations to produce controllable solutions. 

Apart from the nonlinear interactions, by considering the linear coupling effects, 
one can also verify that changes in the SOC parameter are more effective than the 
changes in the Rabi coupling when considering the modifications of the MI band
structures. Besides that, the growth of the verified instability regions is 
larger for the Rabi than for the SOC.
These outcomes can be traced from the results presented in the four panels of Fig.~\ref{f08} 
for $\xi_-$, as one goes from panel (d) [no linear couplings] to panel (b) [only SOC] or 
to panel (c) [only Rabi coupling]. The SO coupling is responsible to produce more 
stable valleys (with corresponding unstable peaks) than the Rabi coupling.
The parametrized regions presented in Figs.~\ref{f03} [no SOC] and ~\ref{f04}
[no Rabi coupling] are also quite indicative in this regard.

{\color{black}
Summarizing the present theoretical work, we have predicted several bands with unstable 
frequencies, by covering a wide range of nonlinear cubic and quintic parameters, together with spin-orbit 
and Rabi couplings. In this regard, our purpose is to be helpful to possible experimental studies,
similar to the investigations that have been performed in Refs.~\cite{2017Nguyen,2017Everitt},
for cigar-type condensed atomic clouds. Our results can already provide an indication of 
appropriate parametrizations which can enhance or reduce MI, 
in possible controllable experimental investigations with different kinds of 
coupled BEC systems.
Even considering that our predictions of MI were limited to cigar-type coupled BEC systems with
SO and Rabi couplings, the actual results obtained of MI emerging from the cubic and quintic 
nonlinearities with the linear couplings are expected to be helpful in other 
experimental setups to study MI, such as the one reported recently in 
Ref.~\cite{2021Vanderhaegen}, which is concerned with MI in 
optics and hydrodynamics.
Finally, within a perspective of further developments of the present investigation, it 
will be interesting to verify how quantum fluctuation can affect the 
MI-parametrized region domains. Another direct extension is to study MI with  
other different aspect ratios for trap confinement.
}

\section*{Acknowledgements}
\noindent SS and LT acknowledges the Funda\c c\~ao de Amparo 
\`a Pesquisa do Estado de S\~ao Paulo (FAPESP) [Contracts No. 2020/02185-1 and No. 2017/05660-0]. 
LT also acknowledges partial support from Conselho Nacional de Desenvolvimento 
Cient\'\i fico e Tecnol\'ogico (CNPq) (Proc. 304469-2019-0).

\clearpage
\onecolumn

\section*{Supplementary material: Numerical simulation} 
\label{sec:appendix}

\begin{figure*}[!ht]
\begin{center}
\includegraphics[width=6.0cm,height=5.9cm]{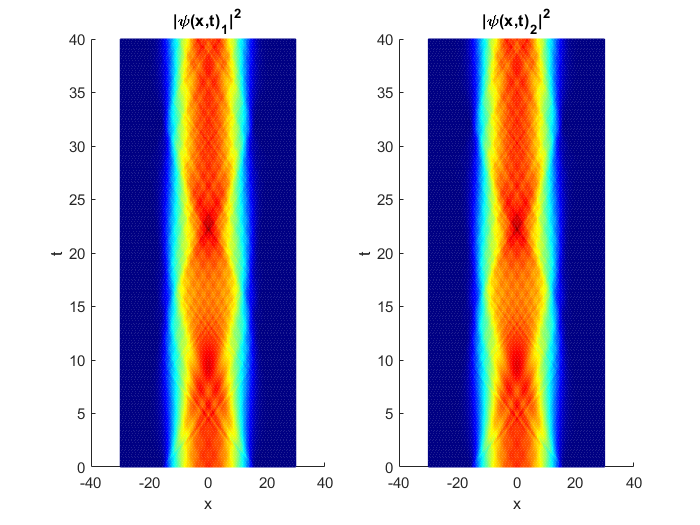}
\includegraphics[width=6.0cm,height=5.9cm]{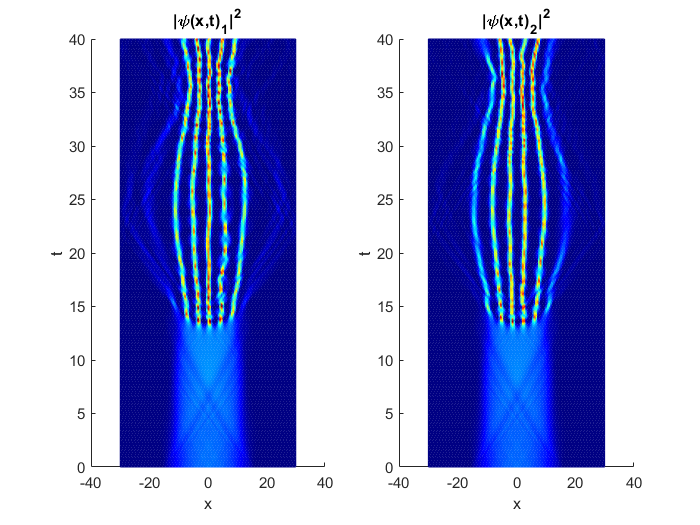}
\includegraphics[width=6.0cm,height=5.9cm]{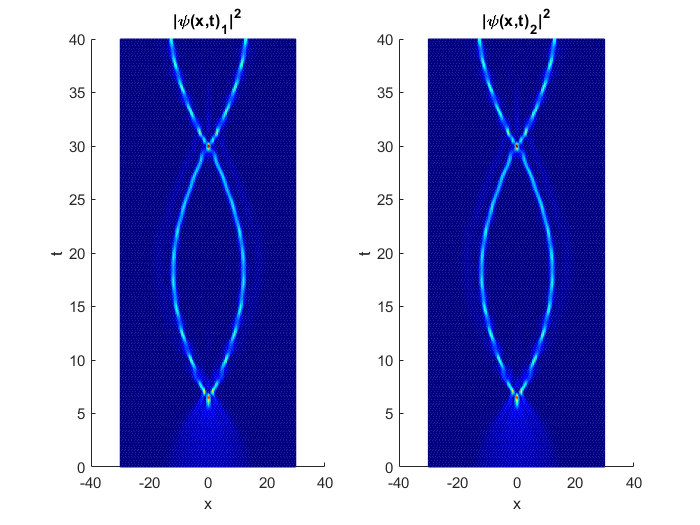}
\end{center}
\vspace{-.5cm}
\caption{(Color online) Time evolutions of the two-component
coupled densities $|\psi_j(x,t)|^2$,
obtained by solving the \eqref{eq:gp1d}, assuming the 
parameters 
$\bar{g}_{ij}$ and 
$\bar{\chi}_{ij}$,
such that  $(\alpha,\beta)=$(1,0), (-1,0), (-1,1.5) [respectively, left, middle, 
and right set of coupled panels], for $\gamma=0,\nu_R=0$.}
\label{f09}
\end{figure*}

\begin{figure*}[!ht]
\begin{center}
\includegraphics[width=6.0cm,height=5.9cm]{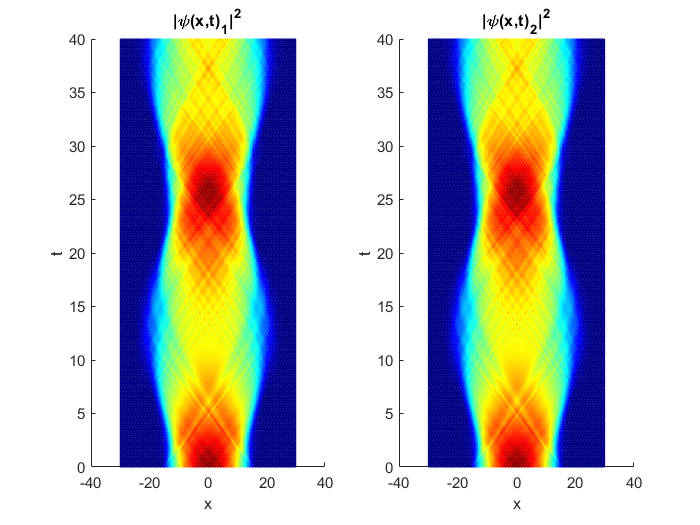}
\includegraphics[width=6.0cm,height=5.9cm]{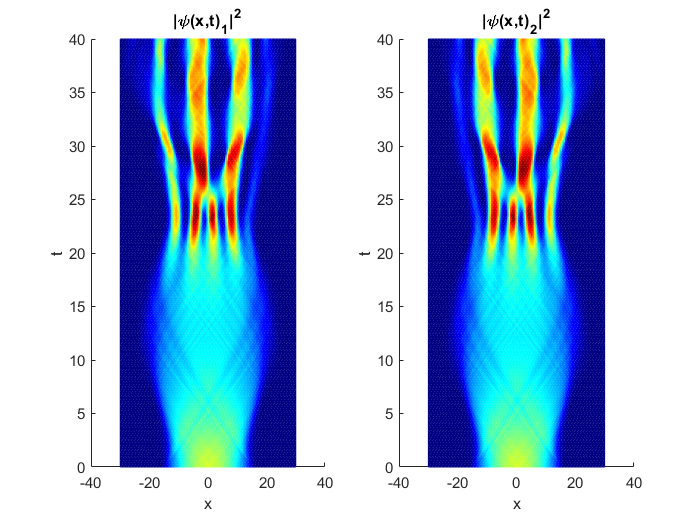}
\includegraphics[width=6.0cm,height=5.9cm]{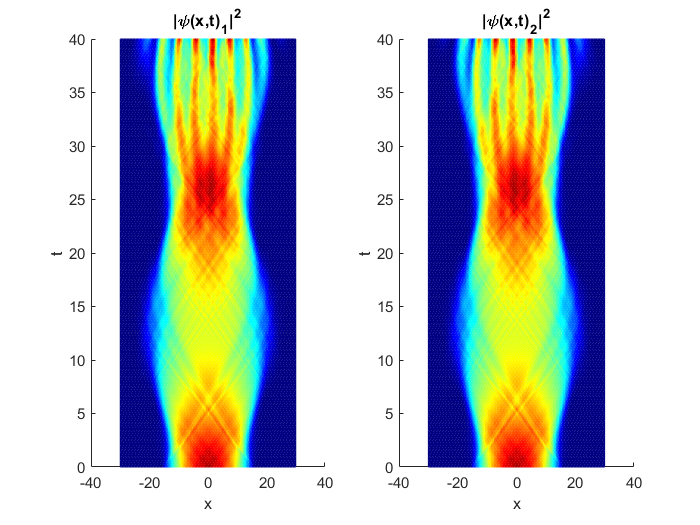}
\end{center}
\vspace{-.5cm}
\caption{(Color online) Time evolutions of the two-component
coupled densities $|\psi_j(x,t)|^2$,
obtained by solving the \eqref{eq:gp1d}, assuming the 
linear couplings have different values, as ($\gamma,\nu_R$)=(0,0), 
(0.001,0), (0.001,0.5). In these cases, we consider the cubic
and quintic parameters fixed, with $\alpha=$0.5 and $\beta=$1.5 in all the panels. }
\label{f10}
\end{figure*}

\noindent The results obtained by the MI analyses are shown to be consistent with previously known 
results in particular cases without quintic nonlinearity, or without linear couplings. 
In addition, with this appendix, we are providing
further support with full-numerical solutions of Eq.~\eqref{eq:gp1d}, exemplified
by  two particular cases with results given in Figs.~\ref{f09} and ~\ref{f10}. 
In Fig.~\ref{f09}, we are presenting three sets of panels for the time evolution of the 
densities $|\psi_j(x,t)|^2$ obtained for the two components, by assuming the parameters 
$(\alpha,\beta)=$(1,0), (-1,0), (-1,1.5) [respectively, left, middle, and right 
set of coupled panels], without linear couplings, which are referring to the 
stability analyses performed in panel (a) of Fig.~\ref{f02}. These results are
shown clearly the transition from stable results (left set of panels) to
unstable ones (right set of panels), in which the middle panels are 
already inside an unstable region, but close to the limit.
In Fig.\ref{f10}, the time evolutions of the densities are included for the case 
in which the linear couplings have different values, as ($\gamma,\nu_R$)=(0,0), 
(0.001,0), (0.001,0.5), with fixed $\alpha=$0.5 and $\beta=$1.5 in all the panels.



\begin{thebibliography}{30}
\bibitem{1967Benjamin} T. B. Benjamin and J. F. Feir, 
{ The disintegration of wavetrains in deep water},
J. Fluid. Mech. \textbf{27} 417 (1967).

\bibitem{1970Hasegawa}A. Hasegawa, 
{ Observation of Self-Trapping Instability of a Plasma Cyclotron Wave in a Computer Experiment},
Phys. Rev. Lett. \textbf{24}, 1165 (1970).

\bibitem{2021Vanderhaegen} G. Vanderhaegen, C. Naveau, P. Szriftgiser, A. Kudlinski, M. Conforti, 
A. Mussot, M. Onorato, S. Trillo, A. Chabchoub, and N. Akhmediev, 
{ ``Extraordinary" modulation instability in optics and hydrodynamics},
PNAS {\bf 118}, e2019348118 (2021).

\bibitem{2020Conforti} M. Conforti, A. Mussot, A. Kudlinski, S. Trillo, N. Akhmediev, 
{ Doubly periodic solutions of the focusing nonlinear Schr\"odinger equation: Recurrence, period doubling, and 
amplification outside the conventional modulation-instability band},
Phys. Rev. A {\bf 101}, 023843 (2020).

{\color{black}
\bibitem{2017Nguyen} J. H. V. Nguyen, D. Luo, R. G. Hulet, 
{ Formation of matter-wave soliton trains by modulational instability},
Science \textbf{356}, 422 (2017).

\bibitem{2017Everitt} P. V. Everitt, 
 M. A. Sooriyabandara, M. Guasoni, P. B. Wigley, C. H. Wei, G. D. McDonald, K. S. Hardman, 
 P. Manju, J. D. Close, C. C. N. Kuhn, S. S. Szigeti, Y. S. Kivshar, and N. P. Robins, 
{Observation of a modulational instability in Bose-Einstein condensates},
Phys. Rev. A {\bf 96}, 041601(R) (2017).
}

\bibitem{2002Khawaja} U. Al Khawaja, H. T. C. Stoof, R. G. Hulet, K.E. Strecker, and G. B. Partridge, 
{ Bright Soliton Trains of Trapped Bose-Einstein Condensates},
Phys. Rev. Lett. \textbf{89}, 200404 (2002).

\bibitem{2002Strecker} K.E. Strecker, G. B. Partridge, A. G. Truscott, and R. G. Hulet, 
{ Formation and propagation of matter-wave soliton trains},
Nature (London) \textbf{417}, 150 (2002).

\bibitem{2004Carr} L. D. Carr and J. Brand, 
{ Spontaneous Soliton Formation and Modulational Instability in Bose-Einstein Condensates},
Phys. Rev. Lett. \textbf{92}, 040401 (2004).

\bibitem{2007Ndzana} F. II Ndzana, A. Mohamadou, T. C. Kofan\'e, 
{ Modulational instability in the cubic-quintic nonlinear Schr\"odinger equation 
through the variational approach},
Opt, Comm. {\bf 275}, 421 (2007).

{\color{black}
\bibitem{Sabari2020} S. Sabari, O.T. Lekeufack, R. Radha, T.C. Kofane, { Interplay of three-body and higher-order interactions on the modulational instability of Bose-Einstein condensate}, J. Opt. Soc. Am. B \textbf{37}, A54 (2020).}

\bibitem{2013Sabari} S. Sabari, E. Wamba, K. Porsezian, A. Mohamadou, and T. C. Kofan\'e, 
{ A variational approach to the modulational-oscillatory instability of 
Bose-Einstein condensates in an optical potential}
Phys. Lett. A \textbf{377}, 2408 (2013).

\bibitem{2015Sabari} S. Sabari, K. Porsezian, and R. Murali, { Modulational and 
oscillatory instabilities of Bose-Einstein condensates with two- and three-body 
interactions trapped in an optical lattice potential}, Phys. Lett. A \textbf{379}, 299 (2015).

\bibitem{2014Wamba} E. Wamba, S. Sabari, K. Porsezian, A. Mohamadou, and T. C. Kofan\'e, 
{ Dynamical instability of a Bose-Einstein condensate with higher-order interactions 
in an optical potential through a variational approach}
Phys. Rev. E  \textbf{89}, 052917 (2014).

{\color{black}\bibitem{Sabari2022} S. Sabari, O.T. Lekeufack, S.B. Yamgoue, R. Tamilthiruvalluvar, R. Radha, { Role of higher-order interactions on the modulational
instability of Bose-Einstein condensate trapped in a periodic
optical lattice}, Int. J. Theor. Phys. \textbf{61}, 222 (2022).
}
\bibitem{Goldstein} E. V. Goldstein and P. Meystre, 
{ Quasiparticle instabilities in multicomponent atomic condensates},
Phys. Rev. A  \textbf{55}, 2935 (1997).

\bibitem{Kasamatsu2004} K. Kasamatsu and M. Tsubota, { Multiple Domain Formation Induced by Modulation Instability in Two-Component Bose-Einstein Condensates}, 
Phys. Rev. Lett. {\bf 93}, 100402 (2004).

\bibitem{Kasamatsu2006} K. Kasamatsu and M. Tsubota, { Modulation instability and 
solitary-wave formation in two-component Bose-Einstein condensates}, Phys. Rev. A  {\bf 74}, 013617 (2006).

{\color{black}
\bibitem{Sabari2019} R. Tamilthiruvalluvar, E. Wamba, S. Sabari, and K. Porsezian, { Impact of higher-order nonlinearity on modulational instability in two-component Bose-Einstein condensates}, Phys. Rev. E \textbf{99}, 032202 (2019).
}
\bibitem{ZRaptiMI} Z. Rapti, A. Trombettoni, P.G. Kevrekidis, D.J. Frantzeskakis, B.A. Malomed, 
A.R. Bishop, { Modulational instabilities and domain walls in coupled discrete nonlinear Schr\"odinger 
equations}, Phys. Lett. A. \textbf{330}, 95 (2004).



\bibitem{1957LHY} T. D. Lee, K. Huang, and C.  Yang, 
{ Eigenvalues and Eigenfunctions of a Bose System of Hard Spheres and Its Low-Temperature Properties},
Phys. Rev. {\bf 106}, 1135 (1957).

\bibitem{2019-Abdullaev} F. Kh. Abdullaev,  A. Gammal, R. K. Kumar, and L. Tomio, 
{ Faraday waves and droplets in quasi-one-dimensional Bose gas mixtures},
J. Phys. B: At. Mol. Opt. Phys. {\bf 52}, 195301 (2019).

\bibitem{2022-Otajonov} S. R. Otajonov, E. N. Tsoy, and F. Kh. Abdullaev,
Modulational instability and quantum droplets in a two-dimensional Bose-Einstein condensate,
Phys. Rev. A {\bf 106}, 033309 (2022).

\bibitem{2015Bhat} I.A. Bhat, T. Mithun, B.A. Malomed, and K. Porsezian, { Modulational 
instability in binary spin-orbit-coupled Bose-Einstein condensates}, Phys. Rev. A  {\bf 92}, 063606 (2015).

\bibitem{2019Mithun} T. Mithun and K. Kasamatsu, { Modulation instability associated 
nonlinear dynamics of spin-orbit coupled Bose-Einstein condensates}, 
J. Phys. B: At. Mol. Opt. Phys. \textbf{52}, 045301 (2019).

{\color{black}\bibitem{Sabari2021} S. Sabari, R. Tamilthiruvalluvar, R. Radha, { Modulational instability of spin-orbit coupled Bose-Einstein
condensates in discrete media}, Phys. Lett. A \textbf{418}, 127696 (2021).}

\bibitem{Lin2011} Y.-J. Lin, K. Jim\'enez-Garc\'{\i}a and I. B. Spielman, 
{ Spin-orbit coupled Bose-Einstein condensates}, 
Nature {\bf  471}, 83 (2011).

\bibitem{Ruseckas2005} J. Ruseckas, G. Juzeli\={u}nas, P. \"{O}hberg, and M. Fleischhauer, 
{ Non-Abelian Gauge Potentials for Ultracold Atoms with Degenerate Dark States}, 
Phys. Rev. Lett. {\bf 95}, 010404 (2005).

\bibitem{Zhang2012} Y. Zhang, L. Mao and C. Zhang, { Mean-field dynamics of spin-orbit 
coupled Bose-Einstein condensates}, 
Phys. Rev. Lett. {\bf  108}, 035302 (2012).

\bibitem{Zutic}I. Zutic, J. Fabian, S.D. Sarma, { Spintronics: fundamentals and applications}, 
Rev. Mod. Phys.  {\bf 76}, 323  (2004).

\bibitem{Nagaosa}N. Nagaosa, J. Sinova, S. Onoda, A.H. MacDonald, and N.P. Ong, 
{ Anomalous hall effect}, 
Rev. Mod. Phys. {\bf 82}, 1539 (2010).

\bibitem{Sau} J.D. Sau, R.M. Lutchyn, S. Tewari, S. Das Sarma, 
{ Generic new platform for topological quantum computation Using Semiconductor heterostructures}, 
Phys. Rev. Lett. {\bf 104}, 040502 (2010).

\bibitem{Mishchenko} E.G. Mishchenko, A.V. Shytov, B.I. Halperin, { Spin current and 
polarization in impure two-dimensional electron systems with spin-orbit coupling}, 
Phys. Rev. Lett. {\bf 93}, 226602 (2004).

\bibitem{Avsar} A. Avsar, J.Y. Tan, T. Taychatanapat, J. Balakrishnan, G.K.W. Koon, Y. Yeo, J. Lahiri, 
A. Carvalho, A.S. Rodin, E.C.T. OFarrell, G. Eda, A.H. Castro Neto, B. Ozyilmaz, 
{ Spin-orbit proximity effect in graphene}, 
Nat. Commun. {\bf 5}, 4875 (2014).

\bibitem{Wang2010} C. Wang, C. Gao, C.M. Jian, and H. Zhai, 
{ Spin-Orbit Coupled Spinor Bose-Einstein Condensates}, 
Phys. Rev. Lett.  {\bf 105}, 160403 (2010).

\bibitem{Li2012} Y. Li, L. P. Pitaevskii, and S. Stringari, 
{ Quantum Tricriticality and Phase Transitions in Spin-Orbit Coupled Bose-Einstein Condensates}, 
Phys. Rev. Lett. {\bf 108}, 225301 (2012).

\bibitem{Achilleos} V. Achilleos, D. J. Frantzeskakis, P.G. Kevrekidis and D.E. Pelinovsky, 
{ Matter-Wave Bright Solitons in Spin-Orbit Coupled Bose-Einstein Condensates}, 
Phys. Rev. Lett. \textbf{110}, 264101 (2013).

\bibitem{Kartashov} Y.V. Kartashov, V.V. Konotop and F.K. Abdullaev, { Gap Solitons in a 
Spin-Orbit-Coupled Bose-Einstein Condensate}, Phys. Rev. Lett. \textbf{111}, 060402 (2013).

\bibitem{Lobanov} V.E. Lobanov, Y.V. Kartashov and V.V. Konotop, { Fundamental, Multipole, 
and Half-Vortex Gap Solitons in Spin-Orbit Coupled Bose-Einstein Condensates}, 
Phys. Rev. Lett. \textbf{112}, 180403 (2014).

\bibitem{Salasnich2013}L. Salasnich and B.A. 
Malomed, {\t Localized modes in dense repulsive and attractive Bose-Einstein condensates 
with spin-orbit and Rabi couplings}, Phys. Rev. A  \textbf{87}, 063625 (2013).

\bibitem{Sakaguchi-Li-Malomed1} H. Sakaguchi, B. Li and B.A. Malomed, { Creation of 
two-dimensional composite solitons in spin-orbit-coupled self-attractive Bose-Einstein 
condensates in free space}, Phys. Rev. E  \textbf{89}, 032920 (2014).

\bibitem{Salasnich-Malomed1} L. Salasnich, W.B. Cardoso and B.A. Malomed, { Localized modes 
in quasi-two-dimensional Bose-Einstein condensates with spin-orbit and Rabi couplings}, 
Phys. Rev. A  \textbf{90}, 033629 (2014).

\bibitem{Sakaguchi-Meqs} H. Sakaguchi and B.A. Malomed, { Discrete and continuum composite 
solitons in Bose-Einstein condensates with the Rashba spin-orbit coupling in one and two 
dimensions}, Phys. Rev. E  \textbf{90},  062922 (2014).

\bibitem{2014-Cheng} Y. Cheng, G. Tang, and S. K. Adhikari, 
{ Localization of a spin-orbit-coupled Bose-Einstein condensate in a bichromatic 
optical lattice}, Phys. Rev. A  {\bf 89}, 063602 (2014).

\bibitem{2016-Salerno} M. Salerno, F. Kh. Abdullaev, A. Gammal, and L. Tomio, 
{ Tunable spin-orbit coupled Bose-Einstein condensates in deep optical lattices},
Phys. Rev. A  {\bf 94}, 043602 (2016).

\bibitem{2018-Abdullaev} F. Kh. Abdullaev, M. Brtka, A. Gammal, and L. Tomio, 
{ Soliton and Josephson-type oscillations in Bose-Einstein condensates with 
spin-orbit coupling and time-varying Raman Frequency},  
Phys. Rev. A   {\bf 97}, 053611 (2018).

\bibitem{levy2007}
S. Levy, E. Lahoud, I. Shomroni, J. Steinhauer, 
{ The a.c. and d.c. Josephson effects in a Bose-Einstein condensate}, 
Nature {\bf 449}, 579 (2007).

\bibitem{abbarchi2013} M. Abbarchi, A. Amo, V.G. Sala, D.D. Solnyshkov, 
H. Flayac, L. Ferrier, I. Sagnes, E. Galopin, A. LemaÃ®tre, G. Malpuech and J. Bloch, 
{ Macroscopic quantum self-trapping and Josephson oscillations of exciton polaritons},
Nature Physics {\bf 9}, 275 (2013).

\bibitem{2021-RavisankarPRA}
R. Ravisankar, H. Fabrelli, A. Gammal, P. Muruganandam, and P. K. Mishra,
{ Effect of Rashba spin-orbit and Rabi couplings on the excitation spectrum
of binary Bose-Einstein condensates},
Phys. Rev. A {\bf 104}, 053315 (2021).

\bibitem{2000Gammal} A. Gammal, T. Frederico, L. Tomio, and P. Chomaz, 
{ Liquid-Gas phase transition in Bose-Einstein Condensates with time evolution}, 
Phys. Rev. A  {\bf 61}, 051602(R) (2000).

\bibitem{2000GammalJPB} A. Gammal, T. Frederico, L. Tomio, and P. Chomaz, 
{ Atomic Bose-Einstein Condensation with Three-Body Interactions and Collective Excitations}. 
Jour. of Phys. B, At. Mol. and Opt. Physics. {\bf 33}, 4053 (2000).

\bibitem{1959Rashba}
E. I. Rashba and V. I. Sheka, 
{ Symmetry of Energy Bands in Crystals of Wurtzite Type II. Symmetry of 
Bands with Spin-Orbit Interaction Included}, 
Fiz. Tverd. Tela: Collected Papers {\bf 2}, 162 (1959).

\bibitem{1955Dresselhaus}
G. Dresselhaus, 
{ Spin-Orbit Coupling effects in Zinc-Blende Structures}, 
Phys. Rev. {\bf 100}, 580 (1955).

\bibitem{1999Braaten} E. Braaten and A. Nieto, 
{ Quantum corrections to the energy density of a homogeneous Bose gas},
Eur. Phys. J. B  {\bf 11}, 143 (1999).

\bibitem{2013Jibbouri} H. Al-Jibbouri, I. Vidanovi\'c, A. Bala\v{z}, and A. Pelster, 
{ Geometric resonances in Bose-Einstein condensates with two-and three-body interactions},
J. Phys. B: At. Mol. Opt. Phys. {\bf 46} 065303 (2013). 

\bibitem{2021Hammond}
A. Hammond, L. Lavoine, and T. Bourdel,  
{ Tunable three-body interactions in driven two-component Bose-Einstein condensates}, 
Phys. Rev. Lett. {\bf 128}, 083401 (2022). 

\bibitem{Agrawal2013} G.P. Agrawal, {Nonlinear Fiber Optics}, 5th edition 
(Academic Press San Diego, 2013).

{\color{black}

\bibitem{2005Abdullaev} F. Kh. Abdullaev,  A. Gammal, A. M. Kamchatnov, and L. Tomio, 
{Dynamics of bright matter-wave solitons in a Bose-Einstein condensate},
Int. J. Mod. Phys. B {\bf 19}, 3415 (2005).

\bibitem{2014Cairncross} W. Cairncross and A. Pelster, Parametric resonance in Bose-Einstein condensates
with periodic modulation of attractive interaction, Eur. Phys. J. D {\bf 68} 106 (2014).
}

\bibitem{Theocharis} G. Theocharis, Z. Rapti, P. G. Kevrekidis, D. J. Frantzeskakis, 
and V. V. Konotop, 
{ Modulational instability of Gross-Pitaevskii-type equations in $1+1$ dimensions}, 
Phys. Rev. A  {\bf 67}, 063610 (2003).

\bibitem{2008Pethick} C. J. Pethick and H. Smith, 
{ Bose-Einstein Condensation in Dilute Gases}, Cambridge University Press, 
Cambridge, UK, 2008.

\bibitem{2005Merhasin} I. M. Merhasin, B. A. Malomed and R. Driben,
{ Transition to miscibility in a binary Bose-Einstein condensate induced by linear coupling},
J. Phys. B: At. Mol. Opt. Phys. {\bf 38} 877 (2005).

\bibitem{Wen2012}L. Wen, w. M. Liu, Y. Cai, J. M. Zhang, and J. Hu, { Controlling phase 
separation of a two-component Bose-Einstein condensate by confinement},
Phys. Rev. A {\bf 85} 043602 (2012).

\bibitem{Kishor2017} R. K. Kumar, P. Muruganandam, L. Tomio
and A. Gammal, 
{ Miscibility in coupled dipolar and non-dipolar Bose-Einstein condensates}, J. Phys. 
Commun. {\bf 1}, 035012 (2017).

\bibitem{Gutierrez2021}E.M. Gutierrez, G.A. de Oliveira, K.M Farias, V.S. Bagnato and 
P.C.M Castilho, 
{ Miscibility Regimes in a $^{23}$Na $^{39}$K Quantum Mixture},
Appl. Sci. {\bf 11}, 9099 (2021).


\end{thebibliography}
\end{document}